\documentclass[leqno,11pt,twoside]{article}
\usepackage{pliska,amsmath,amssymb,mathrsfs}

\def\bbbr{{\Bbb R}}

\def\bbbz{{\Bbb Z}}
\def\openone{\leavevmode\hbox{\small1\kern-3.3pt\normalsize1}}

\def\ad{{\mbox{ad}}}
\def\tr{\mathrm{tr\,}}

\def\im{\mathrm{Im\,}}

\def\diag{\mbox{diag\,}}

\newtheorem{lemma}{Lemma}

\allowdisplaybreaks

\begin{document}

\kwams{
03.75.Lm, 42.65.Tg, 45.20.Jj}
{
Riemann-Hilbert Problems,  commuting operators, integrable 3-wave interactions}

\head
{1--14
}
{Riemann-Hilbert Problems with canonical normalization
and families of commuting operators.
}
{V. S. Gerdjikov \\
\small\dag\,Institute for Nuclear Research and Nuclear Energy\\
Bulgarian Academy of  Sciences \\
72 Tzarigradsko chaussee, 1784 Sofia, Bulgaria 
}
{RHP and families of commuting operators
}
{V. S. Gerdjikov \qquad \today
}
{00
}
{2012
}

\begin{abstract}
\noindent{\sc Abstract.
We start with a Riemann-Hilbert Problems (RHP) with canonical normalization
whose sewing functions depends on several additional variables.
Using Zakharov-Shabat theorem we are able to construct a family of
ordinary differential operators for which the solution of the RHP is a
common fundamental analytic solution. This family of operators obviously
commute. Thus we are able to construct new classes of integrable
nonlinear evolution equations.}
\end{abstract}

  \section{ Introduction}
The development of the soliton theory revealed an important class of NLEE (nonlinear evolution equations)
that describe special types of wave-wave interactions \cite{CaDem,ZMNP,GVYa*08,DokLeb,KRB,ISK*99,Popi} which play
important role in various fields in physics.

A formal approach to the integrable equations started by Gel'fand and Dickey \cite{GeDi,Harnad}
and developed actively later on (see e.g. \cite{Harnad} and the references therein) is well known. It allows one to construct the Lax
representations for important classes of NLEE such as the dispersionless KP hierarchy but it
disregards the spectral properties of the Lax operators.

The topic quickly attracted  mathematicians from spectral theory,
dynamical systems, Lie algebras, Hamiltonian dynamics,
differential geometry, see \cite{ZMNP,MaZa*79,Za*80,Za1*80,Fer*95,GVYa*08,DokLeb}
and the numerous references therein.
It attracted also a number of physicists because they found
important applications  of these NLEE in fluid mechanics, nonlinear optics, superconductivity, plasma physics etc.
As a result many different approaches for investigating the soliton
equations and constructing their Lax representations,
soliton solutions, integrals of motion, Hamiltonian hierarchies etc. were developed, see
\cite{ZM1,MaZa*79,Corn*78,ZMNP,KRB,Za1*80,DJK1*80}. Of course, it is not possible
in a short paper to list all important references that cover the broad topics mentioned above.

The inverse scattering method has been applied to many physically important  multidimensional
evolution equations  including the $N$-wave equation, Davey-Stewartson, Kadomtsev-Petviashvilli etc.
\cite{Za1*76,DJK1*80,DJK2*80,Za*80,Za1*80,ZM1,Nach*Abl*84}.
They have been treated by nonlocal generalizations of the  Riemann-Hilbert
problem and by the $\bar{\partial}$-method.

In the present paper we propose an alternative approach to the same class of
equations using as a starting point  the Riemann-Hilbert problem (RHP)
\cite{ZaSh*74a,ZaSh*79,Za*80,Za1*80,ZMNP,ZaMaLomi}; the importance of the canonical normalization
of RHP was noticed in \cite{21,18}. Our aim is to show that this allows
one to construct rings of commuting operators and in addition gives a tool to
study their spectral properties.

In Section 2 below we start with some preliminaries concerning the RHP. In Section 3 we use the solutions
of the RHP to construct family of jets of order $k$, in Section 4 we list their simplest reductions.
In the last two Sections we demonstrate  how this construction can be used to solve NLEE in two and
higher dimensional space-times. In Section 5 we use jets of order 1 to reproduce well known results
about the $3$-wave equations in two- and three-dimensional space-times. We also demonstrate the
integrability of $N$-wave type equations in higher dimensional space-times.
In Section 6 we use jets of order 2 which allows us to construct
new types of integrable $N$-wave interactions whose interaction terms contain
quadratic and cubic nonlinearities, as well as $x$-derivatives.
These equations also allow integrable extensions to three-dimensional space-time.
The last Section contains discussion and conclusions.

\section{ RHP with canonical normalization}
Let us formulate the RHP:
\begin{equation}\label{eq:rhp}\begin{split}
 \xi^+(\vec{x},t,\lambda) &=  \xi^-(\vec{x},t,\lambda)  G(\vec{x},t,\lambda) , \qquad \lambda^k\in\bbbr, \qquad
 \lim_{\lambda\to\infty} \xi^+(\vec{x},t,\lambda) =\openone ,
 \end{split}\end{equation}
where $\xi^\pm(\vec{x},t,\lambda) $ take values in the simple Lie group  $\mathfrak{G}$ with
Lie algebra $\mathfrak{g}$. $ \xi^+(\vec{x},t,\lambda)$ (resp. $ \xi^-(\vec{x},t,\lambda)$) is
an analytic functions of $\lambda$ for $\im \lambda^k >0$ (resp. for $\im \lambda^k <0$).
 For simplicity we consider particular type of dependence of the sewing
function $G(\vec{x},t,\lambda) $ on the  auxiliary variables:
\begin{equation}\label{eq:Gxt}\begin{split}
i \frac{\partial G}{ \partial x_s } -\lambda^k [J_s,  G(\vec{x},t,\lambda) ] &=0, \qquad
  i \frac{\partial G}{ \partial t } -\lambda^k [K,  G(\vec{x},t,\lambda) ] =0.
\end{split}\end{equation}
where $k \geq 1$ is a fixed integer and $J_s$ are linearly independent elements of the Cartan subalgebra
$J_s \in \mathfrak{h}\subset \mathfrak{g}$.

The canonical normalization of the RHP means that we can introduce the asymptotic expansion
\begin{equation}\label{eq:xi-as}\begin{split}
 \xi^\pm(\vec{x},t,\lambda) = \exp Q(\vec{x},t,\lambda), \qquad Q(\vec{x},t,\lambda)=\sum_{k=1}^{\infty} Q_k(\vec{x},t) \lambda^{-k} .
\end{split}\end{equation}
Since $\xi^\pm(\vec{x},t,\lambda) $ are group elements then all $Q_k(\vec{x},t)\in \mathfrak{g}$. However,
\begin{equation}\label{eq:33w}\begin{split}
 \mathcal{J}_s(\vec{x},t,\lambda)= \xi^\pm (\vec{x},t,\lambda) J_s \hat{\xi}^\pm(\vec{x},t,\lambda),
 \qquad  \mathcal{K}(\vec{x},t,\lambda)=\xi^\pm (\vec{x},t,\lambda) K \hat{\xi}^\pm(\vec{x},t,\lambda),
\end{split}\end{equation}
belong to the algebra $\mathfrak{g}$ for any $J$ and $K$ from $\mathfrak{g}$. If in addition
$K$ also belongs to the Cartan subalgebra $\mathfrak{h}$, then
\begin{equation}\label{eq:jets}\begin{split}
[\mathcal{J}_s(\vec{x},t,\lambda), \mathcal{K}(\vec{x},t,\lambda)] =0.
\end{split}\end{equation}

An important tool in our considerations plays the well known  Zakharov-Shabat theorem \cite{ZaSh*74a,ZaSh*79} formulated below
\begin{theorem}\label{thm:ZSh}
Let $\xi^\pm (x,t,\lambda)$ be solutions to the RHP (\ref{eq:rhp}) whose sewing function depends on
the auxiliary variables $\vec{x}$ and $t$ via eq. (\ref{eq:Gxt}). Then $\xi^\pm (x,t,\lambda)$ are fundamental solutions
of the following set of differential operators:
\begin{equation}\label{eq:xi-L}\begin{split}
L_s\xi^\pm\equiv & i \frac{\partial \xi^\pm }{ \partial x_s } + U_{s}(\vec{x},t,\lambda) \xi^\pm (\vec{x},t,\lambda) -
\lambda^k [J_s,\xi^\pm(\vec{x},t,\lambda)]=0, \\
M\xi^\pm\equiv & i \frac{\partial \xi^\pm }{ \partial t } + V(\vec{x},t,\lambda) \xi^\pm(\vec{x},t,\lambda) -\lambda^k [K,\xi^\pm(\vec{x},t,\lambda)]=0.
\end{split}\end{equation}

\end{theorem}

\begin{proof}
The proof follows the lines of \cite{ZaSh*74a,ZaSh*79}. We introduce the functions:
\begin{equation}\label{eq:gpm}\begin{split}
g_s^\pm(\vec{x},t,\lambda) &= i \frac{\partial \xi^\pm }{ \partial x_s } \hat{\xi}^\pm(\vec{x},t,\lambda) +\lambda^k
\xi^\pm(\vec{x},t,\lambda) J_s \hat{\xi}^\pm(\vec{x},t,\lambda), \\
g^\pm(\vec{x},t,\lambda) &= i \frac{\partial \xi^\pm }{ \partial t } \hat{\xi}^\pm(\vec{x},t,\lambda) +\lambda^k
\xi^\pm(\vec{x},t,\lambda) K \hat{\xi}^\pm(\vec{x},t,\lambda),
\end{split}\end{equation}
and using (\ref{eq:Gxt}) prove that
\begin{equation}\label{eq:3gs}\begin{split}
g_s^+(\vec{x},t,\lambda) = g_s^-(\vec{x},t,\lambda), \qquad g^+(\vec{x},t,\lambda) = g^-(\vec{x},t,\lambda),
\end{split}\end{equation}
which means that these functions are analytic functions of $\lambda$ in the whole complex $\lambda$-plane.
Next we find that:
\begin{equation}\label{eq:g-as}\begin{split}
\lim_{\lambda\to\infty} g_s^+(\vec{x},t,\lambda) = \lambda^k J_s, \qquad
\lim_{\lambda\to\infty} g^+(\vec{x},t,\lambda) = \lambda^k K.
\end{split}\end{equation}
and make use of Liouville theorem to get
\begin{equation}\label{eq:UV}\begin{split}
g_s^+(\vec{x},t,\lambda)& = g_s^-(\vec{x},t,\lambda)= \lambda^k J_s - \sum_{l=1}^{k} U_{s;l} (\vec{x},t) \lambda^{k-l}, \\
g^+(\vec{x},t,\lambda)& = g^-(\vec{x},t,\lambda) =\lambda^k K - \sum_{l=1}^{k} V_{l} (\vec{x},t) \lambda^{k-l}.
\end{split}\end{equation}
We shall see below that the coefficients $U_{s;l} (\vec{x},t)$ and $V_{l} (\vec{x},t)$ can be expressed
in terms of the asymptotic coefficients $Q_s$ in eq. (\ref{eq:xi-as}).
\end{proof}

\begin{lemma}\label{lem:1}
The set of operators $L_s$ and $M$ commute, i.e. the following set of equations hold:
\begin{equation}\label{eq:LsM}\begin{split}
i \frac{\partial  U_{s} }{ \partial x_j } &- i \frac{\partial  U_{j} }{ \partial x_s }
+ [  U_{s}(\vec{x},t,\lambda) - \lambda^k J_s,  U_{j}(\vec{x},t,\lambda) -\lambda^k J_j]=0, \\
i \frac{\partial  U_{s}}{ \partial t } &- i \frac{\partial  V }{ \partial x_s }
+ [  U_{s}(\vec{x},t,\lambda) - \lambda^k J_s,  V(\vec{x},t,\lambda) -\lambda^k K]=0.
\end{split}\end{equation}
where
\begin{equation}\label{eq:UV'}\begin{split}
U_{s}(\vec{x},t,\lambda) =\sum_{l=1}^{k} U_{s;l} (\vec{x},t) \lambda^{k-l}, \qquad
V(\vec{x},t,\lambda) = \sum_{l=1}^{k} V_{l} (\vec{x},t) \lambda^{k-l}.
\end{split}\end{equation}

\end{lemma}

\begin{proof}
The set of the operators $L_s$ and $M$ (\ref{eq:xi-L}) have a common FAS, i.e. they all must commute.
The eqs. (\ref{eq:LsM}) are an immediate consequence of (\ref{eq:xi-L}).
\end{proof}

\section{Jets of order $k$}

In what follows we will consider the jets of order $k$ of $\mathcal{J}(x,\lambda)$ and $\mathcal{K}(x,\lambda)$, see
(\ref{eq:jets}). We introduce them by:
\begin{equation}\label{eq:Jet1}\begin{split}
 \mathcal{J}_{s}(\vec{x},t,\lambda) &\equiv \left( \lambda^k \xi^\pm (\vec{x},t,\lambda) J_l \hat{\xi}^\pm(\vec{x},t,\lambda) \right)_+
 = \lambda^k J_s -  U_{s} (\vec{x},t, \lambda), \\
 \mathcal{K}(\vec{x},t,\lambda) &\equiv \left( \lambda^k \xi^\pm (\vec{x},t,\lambda) K \hat{\xi}^\pm(\vec{x},t,\lambda) \right)_+
 = \lambda^k K -  V (\vec{x},t, \lambda).
\end{split}\end{equation}
The subscript $+$ used above means that we insert the asymptotic expansions of $\xi^\pm$ and their inverse (\ref{eq:xi-as}) and
cut off the terms with negative powers of $\lambda$.

Obviously $U_s(x) \in \mathfrak{g}$ can be expressed in terms of $Q_s(x)$. In doing this we take into
account (\ref{eq:jets}) and obtain  \cite{Helg}
\begin{equation}\label{eq:xi-ex}\begin{split}
\mathcal{J}_s(\vec{x},t,\lambda) = J_s + \sum^{\infty}_{k=1} \frac{1}{k!} \ad_{Q}^k J_s,  \qquad \mathcal{K}(\vec{x},t,\lambda) =
K + \sum^{\infty}_{k=1} \frac{1}{k!} \ad_{Q}^k K,
\end{split}\end{equation}
and therefore for $U_{s;l}$ we get:
\begin{equation}\label{eq:J_k}\begin{split}
U_{s;1}(\vec{x},t) & = -\ad_{Q_1}J_s, \qquad  U_{s;2}(\vec{x},t)  = -\ad_{Q_2}J_s - \frac{1}{2}\ad_{Q_1}^2 J_s \\
U_{s;3}(\vec{x},t) & = -\ad_{Q_3}J_s - \frac{1}{2}\left( \ad_{Q_2}\ad_{Q_1} +\ad_{Q_1}\ad_{Q_2}\right) J_s - \frac{1}{6} \ad_{Q_1}^3J_s \\
  & \vdots \\
U_{s;k}(\vec{x},t) & = -\ad_{Q_k}J_s - \frac{1}{2} \sum_{s+p=k}^{}\ad_{Q_s}\ad_{Q_p}  J_s \\
 &\qquad - \frac{1}{6} \sum_{s+p+r=k}^{}\ad_{Q_s}\ad_{Q_p}\ad_{Q_r}  J_s  - \cdots - \frac{1}{k!} \ad_{Q_1}^k J_s ,
\end{split}\end{equation}
and similar expressions for $V_l(\vec{x}, t) $ with $J_s$ replaced by $K$.

\section{Reductions of polynomial bundles}

An important tool to construct new integrable NLEE is based on Mikhailov's group of reductions
\cite{Mik1}. Below we will use mainly $\bbbz_2$ and  $\bbbz_N$ with $N>2$ reduction groups.
The basic $\bbbz_2$-examples are as follows:
\begin{equation}\label{eq:T10.1}\begin{aligned}
&\mbox{a)} &\quad A \xi^{+,\dag} (x,t,\epsilon \lambda^*) \hat{A} &= \hat{\xi}^-(x,t,\lambda), &\quad
AQ^\dag (x,t,\epsilon\lambda^*) \hat{A} &=-Q(x,t,\lambda), \\
&\mbox{b)} &\quad B \xi^{+,*} (x,t,\epsilon \lambda^*) \hat{B} &= \xi^-(x,t,\lambda), &\quad
BQ^* (x,t,\epsilon\lambda^*) \hat{B} &=Q(x,t,\lambda), \\
&\mbox{c)} &\quad C \xi^{+,T} (x,t, -\lambda) \hat{C} &= \hat{\xi}^-(x,t,\lambda), &\quad
CQ^\dag (x,t,-\lambda) \hat{C} &=-Q(x,t,\lambda),
\end{aligned}\end{equation}
where $\epsilon^2 =1$ and $A$, $B$ and $C$ are elements of the group $\mathfrak{G}$ such that
$A^2=B^2=C^2=\openone$. As for the $\bbbz_N$-reductions we may have:
\begin{equation}\label{eq:T10.2}\begin{aligned}
D \xi^{\pm} (x,t, \omega \lambda) \hat{D} &= \xi^\pm(x,t,\lambda), &\qquad
DQ (x,t, \omega\lambda) \hat{D} &=Q(x,t,\lambda),
\end{aligned}\end{equation}
where $\omega^N=1$ and $D^N=\openone$.

These relations allow us to introduce algebraic relations between the matrix elements of
$Q(x,t,\lambda)$ which will be automatically compatible with the NLEE. The classes of inequivalent reductions of the $N$-wave
equations related to the low-rank simple Lie algebras are given in \cite{Joro,GGK*01,GIK,GKKV*08,GKV*07,GV*V06}.

\section{ On $N$-wave equations ($k=1$) in  2 and more dimensions}

The integrability of the $N$-wave equations has been well known for several decades
now, \cite{MaZa*79,Za1*76,ZM1,ZMNP,GK*81,KRB,DJK2*80}.  Their Lax representation involves two Lax operators
linear in $\lambda$ which are particular case of (\ref{eq:xi-L}) with $k=1$:
\begin{equation}\label{eq:NwL}\begin{split}
L\xi^\pm\equiv & i \frac{\partial \xi^\pm }{ \partial x } + [J,Q(x,t)] \xi^\pm(\vec{x},t,\lambda) -
\lambda [J,\xi^\pm(\vec{x},t,\lambda)]=0, \\
M\xi^\pm\equiv & i \frac{\partial \xi^\pm }{ \partial t } + [K, Q(x,t)] \xi^\pm(\vec{x},t,\lambda) -\lambda [K,\xi^\pm(\vec{x},t,\lambda)]=0.
\end{split}\end{equation}
The corresponding equations take the form:
\begin{equation}\label{eq:Nwe}\begin{split}
i \left[ J, \frac{\partial Q}{ \partial t }\right] - i \left[ K, \frac{\partial Q}{ \partial x }\right]
- \left[ [J, Q], [K,Q(x,t)]\right] =0
\end{split}\end{equation}
In fact  the construction of the FAS for the operator $L$ (\ref{eq:NwL}) \cite{ZaSh*74a,ZaSh*79,ZM1,ZMNP} was the
important step forward, that demonstrated the importance of the RHP for solving integrable equations.

The most important and nontrivial example of such NLEE is the 3-wave equations in two-dimensional space-time
\cite{ZM1,ZMNP}. The most important and non-trivial case corresponds to $\mathfrak{g}\simeq sl(3)$
\begin{equation}\label{eq:33w0}
 Q(x,t) = \left( \begin{array}{ccc} 0 & u_{1} & u_{3} \\ -v_{1} & 0 & u_{2} \\
-v_{3} & -v_{2} & 0   \end{array}\right), \begin{aligned} \qquad J &= \diag (a_1, a_2, a_3), \\
 K &= \diag (b_1, b_2, b_3),
\end{aligned}\end{equation}
with $\tr J=\tr K=0$ and $a_1>a_2>a_3$. We also impose the reduction (\ref{eq:T10.1}a) with $A=\diag( 1, \epsilon_1, \epsilon_2)$
where $\epsilon_1^2=\epsilon_2^2=1$.
Then the 3-wave equations take the form:
\begin{equation}\label{eq:33w2}\begin{split}
\frac{\partial u_{1}}{ \partial t } &- \frac{ a_1 -a_2}{b_1-b_2} \frac{\partial u_{1}}{ \partial x }
+\kappa \epsilon_1 \epsilon_2  u_{2}^* u_{3} =0, \\
\frac{\partial u_{2}}{ \partial t } &- \frac{ a_2 -a_3}{b_2-b_3} \frac{\partial u_{2}}{ \partial x }
+\kappa \epsilon_1  u_{1}^* u_{3} =0, \\
\frac{\partial u_{3}}{ \partial t } &- \frac{ a_1 -a_3}{b_1-b_3} \frac{\partial u_{3}}{ \partial x }
+\kappa \epsilon_2  u_{1}^* u_{2}^* =0,
\end{split}\end{equation}
where
\begin{equation}\label{eq:kappa}\begin{split}
\kappa =a_1(b_2-b_3)-a_2 (b_1-b_3)+a_3 (b_1-b_2).
\end{split}\end{equation}
Depending on the choice of the reduction and on interrelations between the group velocities
the 3-wave interactions may describe qualitatively different processes: soliton decay and
explosive soliton  instability \cite{ZM1,ZMNP}.

In the case of 3-dimensional space-time we consider $Q$ of the form (\ref{eq:33w0}), but now let
$u_j$ and $v_j$ be functions of $x_1=x$, $x_2 =y$ and $t$. Let also $J_1=J$ and $J_2=I=\diag(c_1,c_2.c_3)$. Now
the corresponding solution of the RHP $\xi^\pm(x,y,t,\lambda)$ will be FAS not only of $L$ and $M$
above, but also of
\begin{equation}\label{eq:NwN}\begin{split}
P\xi^\pm\equiv & i \frac{\partial \xi^\pm }{ \partial y } + [I,Q(x,t)] \xi^+(\vec{x},t,\lambda) -
\lambda [I,\xi^+(\vec{x},t,\lambda)]=0,
\end{split}\end{equation}
and all these three operators will mutually commute, i.e. along with $[L,M]=0$ we will have also
 $[L,P]=0$ and  $[P,M]=0$. As a result $Q(x,y,t)$ will satisfy two more NLEE of the form (\ref{eq:33w2B}).
 Obviously it will satisfy also
\begin{equation}\label{eq:33w2B}\begin{aligned}
 2 \frac{\partial u_{1}}{ \partial t } - \frac{ a_1 -a_2}{b_1-b_2} \frac{\partial u_{1}}{ \partial x } -
 \frac{ a_1 -a_2}{c_1-c_2} \frac{\partial u_{1}}{ \partial y } +(\kappa_{1} +\kappa_2 )\epsilon_1 \epsilon_2  u_{2}^* u_{3} =0, \\
2 \frac{\partial u_{2}}{ \partial t } - \frac{ a_1 -a_3}{b_1-b_3} \frac{\partial u_{2}}{ \partial x } -
 \frac{ a_1 -a_3}{c_1-c_3} \frac{\partial u_{2}}{ \partial y } +(\kappa_{1} +\kappa_2 )\epsilon_1  u_{1}^* u_{3} =0, \\
2 \frac{\partial u_{3}}{ \partial t } -  \frac{ a_2-a_3}{b_2-b_3} \frac{\partial u_{3}}{ \partial x } - \frac{ a_2
-a_3}{c_2-c_3} \frac{\partial u_{3}}{\partial y } +(\kappa_{1} +\kappa_2 )\epsilon_2  u_{1}^* u_{2}^* =0.
\end{aligned}\end{equation}
which is linear combination of the three equations mentioned above. Here $\kappa_1=\kappa$ (see eq. (\ref{eq:kappa}) and
\begin{equation}\label{eq:kappa2}\begin{split}
\kappa_2 =a_1(c_2-c_3)-a_2 (c_1-c_3)+a_3 (c_1-c_2).
\end{split}\end{equation}

These three wave equations are related to the  real forms of the algebra $sl(3)$ which has
rank 2. Therefore,  trying to add  more auxiliary variables to the solution of the RHP
will not be effective since only two elements of all $ x_s J_s$ will be linearly
independent.

For $N$-wave equations related to Lie algebras $\mathfrak{g}$ of higher rank $r$ we can add
up to $r$ auxiliary variables. The corresponding PDE takes the form:
\begin{equation}\label{eq:33w1}\begin{split}
r  \frac{\partial Q}{ \partial t }- \sum_{s=1}^{r} (\text{ad}_{J_s}^{-1} \text{ad}_{J})\frac{\partial Q}{ \partial x_s }
 - i\sum_{s=1}^{r}\text{ad}_{J_s}^{-1}   \left[ [J, Q], [J_s ,Q(\vec{x},t)] \right]=0
\end{split}\end{equation}
where $Q$ is an $n\times n$ off-diagonal matrix depending on $r+1$ variables. We remind that
if $J=\diag (a_1,\dots , a_n)$ then
\[ (\ad_{J} Q)_{jk} \equiv ([J,Q])_{jk} = (a_j-a_k)Q_{jk}, \qquad (\ad_{J}^{-1} Q)_{jk}  = \frac{1 }{a_j-a_k}Q_{jk}, \]
and similarly for the other $J_s$.
The  coefficient $r$ multiplying the $t$-derivative can be removed by rescaling of $t$.

Again we can use additional reductions of the type (\ref{eq:T10.1}). More details about these equations
will be given elsewhere.

\section{ New $N$-wave equations ($k=2$) in  2 and more dimensions}
Here we shall give  examples of new types of  $N$-wave equations.
Let us choose again $\mathfrak{g}=sl(3)$. The general form of the potentials is given by
\begin{equation}\label{eq:PQgen}\begin{split}
 Q_1(\vec{x},t) & = \left(\begin{array}{ccc} 0 & u_1 & u_3 \\ -v_1 & 0 & u_2 \\ -v_3 & -v_2 & 0\end{array}\right), \qquad
Q_2(\vec{x},t)  = \left(\begin{array}{ccc} q_{11} & w_1 & w_3 \\ -z_1 & q_{22} & w_2 \\ -z_3 & -z_2 & q_{33}\end{array}\right), \\
\end{split}\end{equation}
We also fix up $k=2$. Then the Lax pair becomes
\begin{equation}\label{eq:lax}\begin{aligned}
L\xi^\pm &\equiv  i \frac{\partial  \xi^\pm}{ \partial x } + U(x,t,\lambda)\xi^\pm (x,t,\lambda) -\lambda^2 ]J,\xi^\pm (x,t,\lambda)]=0, \\
M\xi^\pm &\equiv  i \frac{\partial  \xi^\pm}{ \partial t } + V(x,t,\lambda)\xi^\pm(x,t,\lambda) -\lambda^2 ]K,\xi^\pm (x,t,\lambda)]=0,
\end{aligned}\end{equation}
where using eq. (\ref{eq:J_k})
\begin{equation*}\label{eq:laxUV}\begin{aligned}
U & \equiv U_2 +\lambda U_1 = \left( [J,Q_2(x)] - \frac{1}{2} [[J,Q_1],Q_1(x)] \right) + \lambda [J,Q_1], \\
V& \equiv V_2 +\lambda V_1 = \left( [K,Q_2(x)] - \frac{1}{2} [[K,Q_1],Q_1(x)] \right) + \lambda [K,Q_1].
\end{aligned}\end{equation*}
Note, that this Lax pair is independent of the diagonal elements of $Q_2$.

If we retain the generic potentials (\ref{eq:PQgen}) the Lax pair above will provide us with
a set of 6 new complicated equations for the 6 independent functions $u_j$ and $w_j$. To make the things more
simple we  impose a $\bbbz_2$-reduction of the form (\ref{eq:T10.1}a) with $A=\diag(1,\epsilon,1)$, $\epsilon^2=1$.
Thus $Q_1$ and $Q_2$ get reduced into:
\begin{equation}\label{eq:Q12}\begin{aligned}
Q_1 &=\left(\begin{array}{ccc} 0 & u_1 &0 \\ \epsilon u_1^* & 0 & u_2 \\ 0 & \epsilon u_2^* & 0 \end{array}\right), &\qquad
Q_2 &=\left(\begin{array}{ccc} 0 & 0 & w_3  \\ 0 & 0 & 0 \\ w_3^* & 0 & 0 \end{array}\right),
\end{aligned}\end{equation}
and $J$ and $K$ are as in (\ref{eq:33w0}). Now $L$ and $M$ involve only 3 independent functions.
Skipping the details we get a new type of integrable 3-wave equations:
\begin{equation}\label{eq:n3w}\begin{split}
& i(a_1-a_2) \frac{\partial u_1}{ \partial t } -i(b_1-b_2) \frac{\partial u_1}{ \partial x } +\epsilon \kappa u_2^* u_3
+ \epsilon\frac{\kappa (a_1 - a_2)}{(a_1 - a_3)} u_1|u_2|^2  =0, \\
&i(a_2-a_3) \frac{\partial u_2}{ \partial t } -i(b_2-b_3) \frac{\partial u_2}{ \partial x } +\epsilon \kappa u_1^* u_3
-\epsilon \frac{\kappa (a_2 - a_3)}{(a_1 - a_3)} |u_1|^2 u_2 =0, \\
& i(a_1-a_3) \frac{\partial u_3}{ \partial t } -i(b_1-b_3) \frac{\partial u_3}{ \partial x }
- \frac{i\kappa}{ a_1 -a_3} \frac{\partial (u_1 u_2)}{ \partial x } \\ & \; +\epsilon
\kappa \left(\frac{a_1-a_2}{a_1 -a_3} |u_1|^2 + \frac{a_2-a_3}{a_1 -a_3}  |u_2|^2 \right) u_1u_2  +\epsilon \kappa u_3( |u_1|^2 -|u_2|^2) =0,
\end{split}\end{equation}
where the interaction constant $\kappa$ is given by (\ref{eq:kappa}) and:
\begin{equation}\label{eq:kap}\begin{split}
 u_3 &=w_3 +\frac{ 2a_2 -a_1-a_3}{2(a_1-a_3)} u_1u_2.
\end{split}\end{equation}

The diagonal terms in the Lax representation are $\lambda$-independent. Two of them read:
\begin{equation}\label{eq:3wd}\begin{split}
 i(a_1-a_2) \frac{\partial |u_1|^2}{ \partial t } &-i(b_1-b_2) \frac{\partial |u_1|^2}{ \partial x } -\epsilon \kappa
(u_1u_2 u_3^* - u_1^*u_2^* u_3)  =0, \\
 i(a_2-a_3) \frac{\partial |u_2|^2}{ \partial t } &-i(b_2-b_3) \frac{\partial |u_2|^2}{ \partial x } -\epsilon \kappa
(u_1u_2 u_3^* - u_1^*u_2^* u_3)  =0,
\end{split}\end{equation}
These relations are satisfied identically as a consequence of the NLEE (\ref{eq:n3w}). The third one also vanishes
since $\tr [L,M]=0$.

Let us now consider the case when the sewing function $G$ of the RHP depends on 3 variables: $t$, $x_1=x$ and $x_2=y$
with $J_1=J$ and $J_2=I=\diag(c_1,c_2,c_3)$. For $k=2$ we obtain a set of three ordinary differential operators:
$L$, $M$ (\ref{eq:lax}) and
\begin{equation}\label{eq:lax2}\begin{aligned}
P\xi^\pm &\equiv  i \frac{\partial  \xi^\pm}{ \partial y } + W(x,y,t,\lambda)\xi^\pm (x,y,t,\lambda)-\lambda^2 ]I,\xi^\pm (x,y,t,\lambda)]=0, \\
W & \equiv W_2 +\lambda W_1 \\ &= \left( [I,Q_2(x,y,t)] - \frac{1}{2} [[I,Q_1],Q_1(x,y,t)] \right) + \lambda [I,Q_1(x,y,t)],
\end{aligned}\end{equation}
commuting identically with respect to $\lambda$. It is obvious that $[L,P]=0$ if
\begin{equation}\label{eq:n3w3}\begin{split}
& i(a_1-a_2) \frac{\partial u_1}{ \partial t } -i(c_1-c_2) \frac{\partial u_1}{ \partial y } +\epsilon \kappa_2  u_2^* u_3
+ \epsilon\frac{\kappa_2  (a_1 - a_2)}{(a_1 - a_3)} u_1|u_2|^2  =0, \\
&i(a_2-a_3) \frac{\partial u_2}{ \partial t } -i(c_2-c_3) \frac{\partial u_2}{ \partial y} +\epsilon \kappa_2  u_1^* u_3
-\epsilon \frac{\kappa_2  (a_2 - a_3)}{(a_1 - a_3)} |u_1|^2 u_2 =0, \\
& i(a_1-a_3) \frac{\partial u_3}{ \partial t } -i(c_1-c_3) \frac{\partial u_3}{ \partial y }
- \frac{i\kappa_2 }{ a_1 -a_3} \frac{\partial (u_1 u_2)}{ \partial y } \\ & \; +\epsilon
\kappa_2  \left(\frac{a_1-a_2}{a_1 -a_3} |u_1|^2 + \frac{a_2-a_3}{a_1 -a_3}  |u_2|^2 \right) u_1u_2  +\epsilon \kappa_2  u_3( |u_1|^2 -|u_2|^2) =0,
\end{split}\end{equation}
where $\kappa_2$ is given by eq. (\ref{eq:kappa2}).
It is not difficult to write down the third new 3-wave equation  which is a consequence of the
commutation $[M,P]=0$.

Since the three operators $L$, $M$, and $N$ mutually commute,  $u_1$, $u_2$ and
$u_3$ as functions of $x$, $y$ and $t$ should satisfy {\em simultaneously} the three NLEE of the type
(\ref{eq:n3w}). Therefore  they should satisfy also {\em any} NLEE which is obtained as, say
linear combination of the above:
\begin{equation}\label{eq:n3w33}\begin{split}
& 2i \frac{\partial u_1}{ \partial t } -i(\vec{v}_{(1)}\cdot \nabla)  u_1 +\epsilon (\kappa_1+ \kappa_2)
\left( \frac{ u_2^* u_3}{a_1 - a_2}   + \frac{ u_1|u_2|^2 }{(a_1 - a_3)}\right)  =0, \\
& 2i \frac{\partial u_2}{ \partial t } -i(\vec{v}_{(2)}\cdot \nabla)  u_2 +\epsilon (\kappa_1+ \kappa_2)
\left( \frac{ u_1^* u_3}{a_1 - a_3}   - \frac{ u_2|u_1|^2 }{(a_1 - a_3)}\right)  =0, \\
& 2i \frac{\partial u_3}{ \partial t } -i(\vec{v}_{(3)}\cdot \nabla)  u_3 - i\frac{(\vec{\kappa} \cdot \nabla) (u_1 u_2)}{(a_1 - a_3)^2}
 + \frac{ \epsilon (\kappa_1+ \kappa_2) }{a_1 - a_3} (|u_1|^2 -|u_2|^2 )u_3
\\ &\quad + \frac{ \epsilon (\kappa_1+ \kappa_2) }{(a_1 - a_3)^2}
\left(   (a_1 - a_2)|u_1|^2 +(a_2 - a_3)|u_2|^2     \right) u_1u_2 =0.
\end{split}\end{equation}
Here $\nabla = (\partial_x , \partial_y)^T$, the characteristic velocities $\vec{v}_{(j)}$, $j=1,2,3$ and $\vec{\kappa}$ are
two-component vectors given by:
\begin{equation}\label{eq:}\begin{aligned}
\vec{v}_{(1)} &= \frac{ 1}{a_1-a_2} \left(\begin{array}{c} b_1-b_2 \\ c_1-c_2 \end{array}\right),
 &\quad \vec{v}_{(2)} &= \frac{ 1}{a_2-a_3} \left(\begin{array}{c} b_2-b_3 \\ c_2-c_3 \end{array}\right),\\
\vec{v}_{(3)} &= \frac{ 1}{a_1-a_3} \left(\begin{array}{c} b_1-b_3 \\ c_1-c_3 \end{array}\right),
 &\quad \vec{\kappa} &= \left(\begin{array}{c}  \kappa_1 \\ \kappa_2 \end{array}\right),
\end{aligned}\end{equation}
and $\kappa_1 =\kappa$, see eq. (\ref{eq:kappa}).

\section{Discussion and conclusions}
We have proposed a method for constructing families of commuting operators.
Applied to jets of order 1 with $\mathfrak{g}\simeq sl(3)$ this method reproduces the
well known results for the $3$-wave equations in two- and three-dimensional space-times.
It is shown that $N$-wave equations related to Lie algebras of rank $r$ allow integrable extensions
to $r+1$-dimensional space-times. Below we briefly discuss some open problems and generalizations.

Using jets of order 2 gives us the simplest nontrivial examples for new types of integrable 3-wave equation
whose interaction terms contain quadratic and cubic nonlinearities, as well as $x$-derivatives.
These equations also allow integrable extensions to three-dimensional space-time.

It is not difficult to obtain many other new integrable 3- and $N$-wave equations.
Indeed, one can choose: i)  higher rank simple Lie algebras; ii) different types of grading;
iii) different power $k$ of the polynomials $U(\vec{x},t,\lambda)$ and $V(\vec{x},t,\lambda)$
and iv) different reductions of $U$ and $V$.

These new NLEE  must be Hamiltonian. It is natural to view the jets $U(\vec{x},t,\lambda)$  as  elements
of more complicated co-adjoint orbits of the relevant Kac-Moody algebra, generated by the chosen
grading of $\mathfrak{f}$, see \cite{KuRei,Rei,ReiSTSH}.

By construction, the method allows treating  multi-dimensional NLEE. In the
examples above we used the algebra $sl(3)$ and demonstrated integrable $3$-wave equations in
$2+1$-dimensional space-time. If we want  to study new types of integrable $N$-wave models in
$r+1$ space-time dimensions we have to consider Lie algebras of rank $r$ and accordingly larger values for $N$.

The method allows one also to apply Zakharov-Shabat dressing method \cite{ZaSh*74a,ZaSh*79,MiZa*80,RI}
for constructing their explicit ($N$-soliton) solutions.
Instead of solving the inverse scattering problem for $L$ we would rather deal with a Riemann-Hilbert
problem with canonical normalization. For polynomials of order $k$ the contour on which the RHP is
defined consists of $k$ straight lines $l_k \colon \arg \lambda =\pi i/k$ passing through the
origin. Of course, it may necessary to use
dressing factors with more specific $\lambda$-dependence.

This approach can be used also to analyze the NLEE derived by Gel'fand-Dickey approach \cite{GeDi,Harnad}.
It would provide the possibility to systematically construct the spectral decompositions that
linearize the relevant NLEE \cite{IP2,GVYa*08}. Still more challenging is to study the soliton
interactions of the new $N$-wave equations.

\end{document}